\documentstyle[aps,preprint,psfig,tighten]{revtex}

\newcommand{\be}{\begin{equation}}
\newcommand{\ee}{\end{equation}}
\newcommand{\bea}{\begin{eqnarray}}
\newcommand{\eea}{\end{eqnarray}}
\newcommand{\bdm}{\begin{displaymath}}
\newcommand{\edm}{\end{displaymath}}

\newcommand{\cH}{{\cal H}}    

\newcommand{\bR}{{\bf R}}
\newcommand{\bq}{{\bf q}}
\newcommand{\bQ}{{\bf Q}}
\newcommand{\bk}{{\bf k}}
\newcommand{\bG}{{\bf G}}

\newcommand{\bS}{{\bf S}}
\newcommand{\bs}{{\bf s}}

\begin{document}

\bibliographystyle{prsty-all_names}


\title{Theoretical model for the superconducting and magnetically ordered
       borocarbides}
\author{A. Amici$^1$, P. Thalmeier$^2$ and P. Fulde$^1$\\
$^1${\em Max-Planck-Institut f\"{u}r Physik komplexer Systeme, 
D-01187 Dresden, Germany}\\
$^2${\em Max-Planck-Institut f\"{u}r Chemische Physik fester Stoffe, 
D-01187 Dresden, Germany}}

\maketitle

\begin{abstract}
We present a theory of superconductivity
in presence of a general magnetic structure in a form
suitable for the description of complex magnetic
phases encountered in borocarbides.
The theory, complemented with some details of the band structure
and with the magnetic
phase diagram, may explain the nearly reentrant
behaviour and the anisotropy of the upper critical field of HoNi$_2$B$_2$C.
The onset of the helical magnetic order 
depresses superconductivity via
the reduction of the interaction
between phonons and electrons caused  by the formation of
magnetic Bloch states.
At mean field level, no additional suppression of superconductivity is
introduced by the incommensurability of the helical phase.

\vspace{.5cm}
\end{abstract}

In 1957 V.~L.~Ginzburg first showed that superconductivity
and long range ferromagnetic order compete with each other making
their mutual coexistence nearly impossible \cite{MS:gin57}.
On the other hand in 1963 W.~Baltensperger and S.~Str\"{a}ssler 
pointed out that in the case of antiferromagnetic order 
the two order parameters may actually coexist \cite{MS:bal63}.
Two families of magnetically 
ordered superconductors, the rare earth ternary compounds
$R$Rh$_4$B$_4$ and $R$Mo$_6$S$_8$, 
were almost simultaneously discovered in the late seventies and both 
theoretical predictions proved to be correct,
see e.g. Ref. \cite{MS:map82} for an extensive review.

More recently the discovery of quaternary compounds of the family
$R$Ni$_2$B$_2$C, with $R$~=~Lu, Y 
or rare earth element, \cite{bo:nag94,bo:nat94} 
renewed the interest in the issue
because more complex magnetic structures were observed 
to coexist or to compete with superconductivity 
(for a review see Ref. \cite{bo:lyn97}).
For $R$~=~Er and Tm the magnetic structures coexisting with superconductivity
are incommensurate transversely polarised spin-density waves:
$T_c$~=~11~K and $T_N$~=~6.8~K for Er and
$T_c$~=~11~K and $T_N$~=~ 1.5~K for Tm. 
DyNi$_2$B$_2$C, with $T_c$~=~6~K, 
orders in a commensurate antiferromagnetic state
at $T_N$~=~10.6~K and it
is the only borocarbide compound with $T_N>T_c$. 
The case of HoNi$_2$B$_2$C ($T_c$~=~8-9~K) is more complex:
the transitions
into two incommensurate magnetically ordered 
states, at $T_{IC}^c\sim T_{IC}^a\sim$~6~K
with wave vectors $\bQ_c$~=~0.91${\bf c^*}$ and $\bQ_a$~=~0.55${\bf a^*}$
respectively,
coincide with a deep depression of the superconducting 
upper critical field \cite{bo:eis94}, while below
the transition temperature $T_N$~$\sim$~5~K into a commensurate
antiferromagnetic state, with $\bQ_{AF}$~=~${\bf c^*}$,
$H_{c2}$ rapidly recovers. The anisotropic
upper-critical-field phase diagram
with the magnetic field along the symmetry directions
of the crystal is given in Ref. \cite{Ho:kru96}.
Experimental data on pseudo-quaternary borocarbides \cite{bo:cho96,bo:sch97}
add insight on the dramatic effects that the presence of magnetic order
produces on superconductivity. On the other hand
no difference is found in the magnetic properties of related
superconducting and non-superconducting compounds.

Borocarbides have a body-centered-tetragonal lattice structure
with \emph{I4/mmm} space group symmetry.
In spite of their layered structure they
possess three dimensional conduction bands of mainly
Ni 3$d$ character \cite{bo:mat94,bo:pic94}. The strongly 
anisotropic magnetic properties are associated with
the localized 4$f$ electrons of the rare earths.
The two types of electrons interact weakly via
the local spin exchange on the rare earth sites.
Relying on experimental evidence we consider the
4$f$-electrons system to be independent from the
state (normal or superconducting) of the conduction electrons.
Theoretically this is justified by the separation in the
energy scales associated with the two ordering phenomena:
$E_{MO} \sim $ k$_B T_M$ for magnetic order and
$E_{SC} \sim ($k$_B T_c)^2/E_F$ for the superconductivity.
Careful treatment is needed for
magnetic structures with  small $\bq$ (e.g. $|\bq| < 1/\xi$, with $\xi$
the coherence length)
or originating from nesting features of the Fermi surface.
In fact the opening of the small superconducting gap affects significantly
the RKKY-type magnetic interaction in the regions of $\bq$-space close
to zero or to a nesting vector.
However, away from these special $\bq$-points,
the structure of the magnetic interaction 
is related to electron-hole excitations of all energies
and is independent from the presence of superconductivity.
Therefore, it is possible to take the magnetic properties of borocarbides
from experiments or from an independent microscopic magnetic 
model and concentrate on the influence which the molecular field
of the ordered moments has on the conducting electrons.
Similar approaches have been used in the cases of the
antiferromagnetic \cite{MS:mor80,MS:zwi81} and
small $\bq$ helical order \cite{MS:mor81}
coexisting with superconductivity in $R$Rh$_4$B$_4$ and $R$Mo$_6$S$_8$.
More recently 
the same technique has been used for
a qualitative discussion of the properties of HoNi$_2$B$_2$C
\cite{Ho:mor96a,Ho:mor96b}.

The aim of this work is to present a theory of superconductivity 
in an arbitrarily modulated exchange field capable of describing
the magnetic structures encountered in the borocarbides.
Applied to the upper-critical-field phase diagram of HoNi$_2$B$_2$C,
the theory reproduces naturally its several anomalous features
and gives an explanation for its nearly reentrant behaviour.

We introduce the following BCS model Hamiltonian for the conduction
electrons in the presence of a periodic molecular field:
\bea
\cH   &=& \cH_b + \cH_{b\mbox{-}4f} + \cH_{b\mbox{-}b} \label{eq:Htot} \\
\cH_b &=&
  \sum_{k}\epsilon_{\bf k} \, 
  c^{+}_{k}c_{k} \label{eq:Hb}\\
\cH_{b\mbox{-}4f} &=&
  \sum_\bq I(\bq) \, \left< \bS_\bq \right> \cdot \bs_\bq \label{eq:Hint} \\
\cH_{b\mbox{-}b} &=&
  \frac{1}{2}\sum_{k'_1 k'_2 k_2 k_1}V^{ k_2 k_1}_{k'_2 k'_1} \, 
    c^{+}_{{k'_1}}c^{+}_{{k'_2}}
    c_{k_2}c_{k_1}. \label{eq:Hb-b}
\eea
The term $\cH_b$ is the Hamiltonian
for the free conduction band electrons, $c^+_k$
($= c^+_{\bk\sigma}$) and $c_k$ ($= c_{\bk\sigma}$) are respectively the
creation and annihilation operators for a state with
quantum number $k = (\bk,\sigma)$.
The term $\cH_{b\mbox{-}4f}$ is the
exchange interaction between 
the spin of the local 4$f$ moments $\left< \bS_{\bR_i} \right>$
(with Fourier transform $\left< \bS_\bq \right>$)
and the spin of the conduction electrons, 
$
  \bs_\bq = \sum_{{\bf k}\sigma\sigma'}
   c^{+}_{{\bf k+q}\sigma'}
   \mbox{\boldmath $\sigma$}_{\sigma'\sigma} \,
    c_{{\bf k}\sigma}
$.
At the moment we do not specify any magnetic configuration
and we only require the periodicity of the known function
$\left< \bS_{\bR_i} \right>$.
$\cH_{b\mbox{-}b}$ is the intra-band interaction term whose attractive
part leads to superconductivity. 
Note that 
$V^{k_2 k_1}_{k'_2 k'_1}=0$ for all matrix elements
not conserving spin and crystal momentum
(in the form $\bk'_2 + \bk'_1 + \bG = \bk_2 + \bk_1$, with $\bG$ a reciprocal 
lattice vector).

The first two terms of the total Hamiltonian are bilinear
and therefore may be easily diagonalized for the magnetic structures
encountered in the borocarbides (i.e.,
helices and SDWs).
The magnetic Bloch-states obtained this way, 
with creation and annihilation operators
$\tilde{c}^{+}_{k}$
(= $\tilde{c}^{+}_{\bk\tau}$) and $\tilde{c}_{k}$ 
(= $\tilde{c}_{\bk\tau}$),
are labelled by the momentum $\bk$ and the 
quantum number $\tau$ = + or $-$. As a convention we assume
the momentum $\bk$ to belong to the non-magnetic Brillouin zone,
in order not to introduce an additional magnetic band index. 
Correspondingly the
law of crystal momentum conservation is satisfied modulo
a vector in the magnetic reciprocal lattice $\tilde{\bG}$.
The magnetic reciprocal lattice may be constructed
by adding to every non-magnetic vector $\bG$ a finite set
of vectors $\{\bG^M\}$, and  momentum
conservation requires:
$
\bk'_2 + \bk'_1 + \bG + \bG^M= \bk_2 + \bk_1
$.
In the new basis the Hamiltonian (\ref{eq:Htot}) reduces to
$\tilde{\cH} = \tilde{\cH}_b + \tilde{\cH}_{b-b}$
with $\tilde{\cH}_b$ and $\tilde{\cH}_{b-b}$
given by eqs. (\ref{eq:Hb}) and (\ref{eq:Hb-b})
with all the symbols written with tildes. When
spin-degeneracy is lifted the energy $\epsilon_\bk$ becomes $\tilde{\epsilon}_k$.
Now the Hamiltonian $\tilde{\cH}$ is formally very similar to
the usual BCS Hamiltonian.
The main differences are the modified law of the momentum conservation
and the additional $\bk$-dependence of
the magnetic energy bands $\tilde{\epsilon}_k$
and the electron-electron interaction $\tilde{V}^{ k_2 k_1}_{k'_2 k'_1}$.
The mean field approximation may be applied 
to $\tilde{\cH}$
in the same way as in the non-magnetic case,
via the introduction of the gap functions of the new
magnetic eigenstates $\Delta^{\tau'\tau}_{\bG^M}(\bk)$
corresponding to the anomalous Green functions 
$\left<\tilde{c}_{-\bk+\bG^M\tau'} \tilde{c}_{\bk\tau}\right>$.
The function $\Delta^{\tau'\tau}_{\bG^M}(\bk)$ is actually a matrix in the
$\tau$ indices in order to include odd and even parities.

Some general qualitative properties of the magnetic Bloch-states
and the implications of their use for superconductivity have been discussed
in Ref. \cite{bo:ami99}. However,
numerical complications due to the non-trivial
$\bk$-dependence introduced so far prevent 
the solution of the self-consistent gap equations in the general case.
In order to proceed further the explicit form of the
underlying magnetic structure of a particular material is needed.

\begin{figure}
\psfig{file=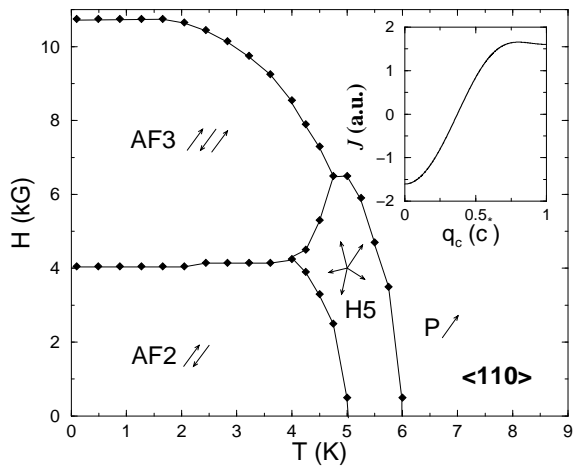,width=7.5cm}
\protect\caption{
Magnetic phase diagram for HoNi$_2$B$_2$C
in the H-T plane obtained from the model discussed in Ref.
\protect\cite{Ho:ami98}. 
The magnetic field lies along the 
$\left<110\right>$ easy direction ($\nearrow$).
The phases have ferromagnetic alignment in the $ab$-plane
and the stacking sequences along the $c$-axis shown in the figure.
Inset: the RKKY interaction function ${\cal J}(\bq)$.}
\label{fig:Mphd}
\end{figure}

The obvious choice for the first application of this formalism
is the analysis of the much debated issue of the
almost reentrant upper critical field in HoNi$_2$B$_2$C.
In what follows we concentrate on the magnetic ordered states
with periodicity along the $c$-axis:
the high temperature incommensurate helix (\bQ$_c$~=~0.91{\bf c}$^*$)
and the low temperature commensurate antiferromagnetic state 
(\bQ$_{AF}$~=~{\bf c}$^*$).
The qualitative features of the lock-in transition
are reproduced by theoretical models including the RKKY interaction
and the crystalline electric field (CEF).
Two such models \cite{Ho:ami98,Ho:pok98}
were first developed to account for the complex
meta-magnetic phase diagram of HoNi$_2$B$_2$C at $T$~=~2K \cite{Ho:can97}. 
In particular the model in Ref. \cite{Ho:ami98},
which includes the actual CEF states of Ho, is capable to
produce a temperature dependent phase diagram, which is used
here as input.
All the parameters of this model have been fitted to the
low temperature magnetic properties of the normal state \cite{Ho:ami98} 
and are not considered adjustable
quantities in what follows.
In fig.~\ref{fig:Mphd} we show
the magnetic phase diagram of the
model as a function of the temperature
and of the magnetic field along the easy axis of the Ho moments,
given by the crystallographic $\left<110\right>$ direction.
Input data for the coexistence analysis are
the following calculated quantities:
the underlying magnetic structure, the temperature dependent 
magnetic order parameter $S(T)$, the magnetic energies and the magnetic
susceptibility.

Given the helical magnetic structure
$\bS_{{\bf R}_i} = S
  [
    {\bf \hat{a}}\cos{(\bQ\cdot {\bf R}_i)}
+
    {\bf \hat{b}}\sin{(\bQ\cdot {\bf R}_i)}
  ]$,
the interaction Hamiltonian in eq. (\ref{eq:Hint})
has the following explicit form:
\be
\cH_{b\mbox{-}4f} =
  IS
    \sum_{{\bf k}\sigma\sigma'}
      \left(
        c^{+}_{{\bf k-Q}\sigma'}
        \sigma^+_{\sigma'\sigma} \,
        c_{{\bf k}\sigma} +
        c^{+}_{{\bf k+Q}\sigma'}
        \sigma^-_{\sigma'\sigma} \,
        c_{{\bf k}\sigma}
     \right) 
\label{eq:Hexplicit}
\ee
with $\sigma^\pm=\sigma_x\pm i\sigma_y$. 
The corresponding magnetic states
are found via simple Bogoliubov transformation
of the type
$
  \tilde{c}^{+}_{{\bf k}+}=u_\bk c^{+}_{{\bf k}\uparrow} 
   +
  v_\bk c_{{\bf k+Q}\downarrow}$
where the state ($\bk$, $\uparrow$) mixes only with one other state,
namely ($\bk$+$\bQ$,$\downarrow$), independently of the value
of $\bQ$~\cite{RE:her66}. 
The Bogoliubov coefficients $u_\bk$ and $v_\bk$, 
the new energies $\tilde{\epsilon}_k$ and the new scattering matrix can
be derived analytically in close analogy with the antiferromagnetic
case. In particular we have:
\be
u^2_{\bf k}-v^2_{\bf k}=
               \sqrt{
                 \frac{(\epsilon_{\bf k}-\epsilon_{\bf k+Q})^2}
                      {(\epsilon_{\bf k}-\epsilon_{\bf k+Q})^2 + 4I^2S^2}
               }
\ee
and:
\be
\tilde{\epsilon}_{{\bf k}\pm} =
  \frac{\epsilon_{\bf k}+\epsilon_{\bf k\pm Q}}{2}+
  \frac{\epsilon_{\bf k}-\epsilon_{\bf k\pm Q}}{2}
  \sqrt{1+\frac{4I^2S^2}{(\epsilon_{\bf k}-\epsilon_{\bf k\pm Q})^2}}
\ee
The expression for $\tilde{V}^{ k_2 k_1}_{k'_2 k'_1}$ 
(not shown) is analytic as well. However,
even assuming the simple BCS interaction potential
in the non-magnetic state, it is rather complicated.
It is important to note
that the magnetic energy bands $\tilde{\epsilon}_k$ possess 
magnetic gaps only
for the two pairs of magnetic-Bragg planes orthogonal to the $c$-axis at
distances Q/2 and ($c^*$ - Q/2) from the $\Gamma$-point and
that all the magnetic quantities differ from the
non-magnetic ones only in narrow regions around these
planes~\cite{RE:her66}. The presence of only four active magnetic-Bragg planes 
for any commensurate value of the ordering vector $\bQ$ 
make the following arguments
hold for structures with arbitrary
periodicity. If we assume that the onset of magnetic order affects the
superconducting state as a perturbation, the
only sizable components of the gap function matrix are
$\Delta^{+-}_{\bf 0}=-\Delta^{ -+}_{\bf 0}\equiv \Delta$ \cite{Ho:mor96b}.
After some algebraic manipulations the gap equation 
simplifies considerably and the gap function
may be written as $\Delta({\bf k},T)
= (u^2_{\bf k}-v^2_{\bf k})\Delta(T)$ \cite{Ho:mor96b}.
This leads to:
\be
\Delta(T)=
\int \limits_{0}^{\omega_D} d\epsilon
\left( V \int \limits_{MFS} \frac{dS'}{(2\pi)^3}
\frac{\left(
u^2_{\bf k'}-v^2_{\bf k'}
\right)^2
}{|\nabla_{\bf k'}\tilde{\epsilon}_{\bf k'}|}\right)
\frac{
\Delta(T)
{\cal F}(T)}{\sqrt{\epsilon^2+\Delta^2(T)}}
\label{eq:self}
\ee
where ${\cal F}(T) \equiv (1- 2n_{\bk'})$ 
takes into account the occupation of the
electronic states and we have approximated $\Delta({\bf k},T)$
with $\Delta(T)$ in the square root. 
Equation (\ref{eq:self})
corresponds to the usual BCS self-consistent equation with
an effective interaction parameter $\lambda_e$(T)
defined as the term into brackets. $\lambda_e$(T) depends on
the underlying magnetic state through the Bogoliubov coefficients
and through the shape of the magnetic Fermi surface (MFS).
Since all the anomalous magnetic $\bk$-dependencies come from 
the regions
where the Fermi surface intersects the four Bragg planes,
the difference $\Delta\lambda(T) = \lambda - \lambda_e(T)$
between the actual electron-phonon interaction parameter ($\lambda$) and
the effective one
may be expanded in terms of $\frac{IS(T)}{\epsilon_F}$.
This simplifies considerably the analysis because only very limited
knowledge of the actual band structure is needed in order to estimate
the variation $\Delta\lambda$(T) of the effective interaction.
Therefore the high temperature interaction parameter $\lambda$
can be considered a phenomenological parameter to be determined
by the transition temperature $T_c$.
An inspection of the LDA band structure of
HoNi$_2$B$_2$C \cite{bo:dre99} suggests approximate rotational symmetry
at the intersection between the Fermi
surface and the Bragg planes corresponding to
the observed magnetic phases. With the assumption of rotational symmetry
the only parameters we need for the band structure are
the radial and vertical component of the Fermi velocity
$v_r$, $v_z$ at the intersections.
To  first order in the parameter $\frac{IS(T)}{\epsilon_F}$, 
for the two pairs of Bragg planes we have:
\be
 \Delta\lambda(T) = -\frac{V}{2\pi\,\hbar^2}
\,\frac{k_r}{v_r v_z}\,IS(T)
\label{eq:deltalambda}
\ee
where $k_r$ is the radius of the intersection.
We note that the two components of the Fermi velocity enter
eq.~(\ref{eq:deltalambda}) independently and the 
perturbation expansion actually breaks down if
$v_r$ or $v_z$ are much smaller than $v_F$.
In particular in the case of nesting we have $v_r\ll v_F$
and $\Delta\lambda(T)$ is not a linear function of S(T).
However the magnetic structures along the $c$-axis
are not linked to nesting and may safely be treated within
our perturbation expansion.
We assume the
relevant phonons to have an average energy $\omega_D$~=~40meV
and with the value $T_c$~=~8.5K, the BCS formula gives
us a value of $\lambda$~=~0.25.
Taking into account the closeness of the two magnetic ordering vectors
Q$_{AF}={\bf c^*}$ and Q$_{c}=0.91{\bf c^*}$ 
we assume $v_r$, $v_z$ and $k_r$ not to change with the lock-in.
This means that we treat the incommensurate and the commensurate
phases exactly on the same footing.

\begin{figure}
\psfig{file=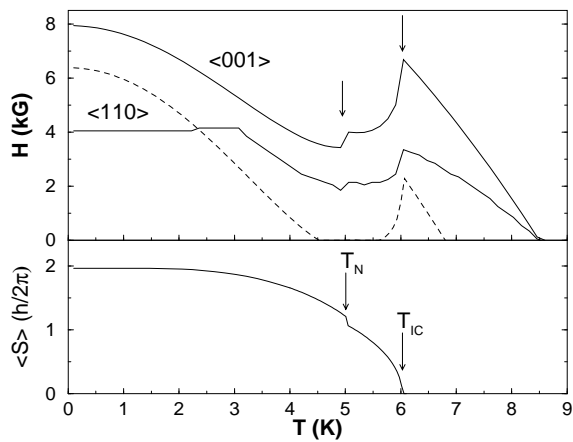,width=7.5cm}
\protect\caption{
Lower panel: size of the magnetic order parameter $\left<S\right>$
vs. temperature. For temperatures above (below) the first order lock-in 
transition temperature $T_N$ the H5 (AF2) order parameter is plotted.
Upper panel: Upper critical field curves H$_{c2}$(T) 
along the $\left<001\right>$ and
along the $\left<110\right>$ directions.
The dashed line the $\left<001\right>$ curve with $T_c$ reduced to 6.8~K}
\label{fig:hc2}
\end{figure}

The main features of the experimental anisotropic phase diagram 
of HoNi$_2$B$_2$C are reported in Ref. \cite{Ho:kru96}
which we will use in order to compare our results.
The upper critical field may be calculated from the equation:
\be
H^{\left<110\right>}_{c2}(T) = B^{BCS}_{c2}(\lambda (T), T)-M(T)
\ee
where $B^{BCS}_{c2}(\lambda (T), T)$ is the critical field value
in the non-magnetic BCS case and $M(T)$ is the
magnetization in the normal state.
Since the magnetic response of HoNi$_2$B$_2$C is very weak along the 
$c$-axis we may neglect the magnetization term 
and the upper critical field 
curve that we obtain is the upper line ($\left<001\right>$)
in fig.~\ref{fig:hc2}
where we have used the value for $\Delta\lambda(0K)/\lambda$~=~0.12.
The depression of the
critical field is then due to the onset of the magnetic order 
altogether and
not to its incommensurate nature.
The depression results from a small but rapid
decrease of the effective interaction parameter $\lambda_e(T)$
related to any helical structure, regardless
to its periodicity. Furthermore the almost reentrant behaviour
shown in fig.~\ref{fig:hc2} is produced by a reduction
in the value of $\lambda$ by only 12\%.
The small jump at the lock-in transition
is due to the discontinuity of the single-site magnetization
going from 5-cell helix to the antiferromagnet shown in the lower
part of fig.~\ref{fig:hc2}.

In order to reproduce the other features of 
the anisotropic upper critical field phase
diagram is important to take into account the strong
anisotropic magnetic response of the system and the presence
of metamagnetic transitions in the same range of fields and temperatures.
In fig.~\ref{fig:hc2} is shown the curve for the external field along the
$\left<110\right>$ direction calculated with the $M(T)$ response
of the magnetic system. 
With respect to the $\left<001\right>$ curve,
the peak around 6K is strongly reduced by the magnetization,
while for temperatures below 4K the two curves approach
each other due to the saturation of the ordered microscopic magnetic
moments and the subsequently reduced magnetic response of the system.
The low temperature plateau is due to the transition
to the metamagnetic phase
AF3  (compare with fig.~\ref{fig:Mphd}) 
which has a large ferromagnetic component and
suppresses superconductivity much more strongly than the low
field antiferromagnet.
The $\left<100\right>$ critical field curve
(not shown) has almost the same shape
as the $\left<110\right>$ curve for temperatures higher than 3K, 
but the plateau is reached at a slightly smaller temperature and for a 
slightly larger value of the magnetic field. This corresponds to the
upper shift of the metamagnetic transition to AF3.
All these features are in quantitative agreement with the
experimental data \cite{Ho:kru96}.

In addition our model explains in a natural way
the fact that HoNi$_2$B$_2$C samples with $T_c$ reduced
via different techniques, i.e. with Co doping \cite{bo:sch97},
actually reenter the normal state in a temperature region around
the lock-in transition $T_N$. The dashed line in fig.~\ref{fig:hc2}
is the upper critical field along the $c$-axis
obtained leaving all the parameters except $T_c$ unchanged.

In conclusion, we derived a theory of superconductivity
in a magnetically ordered background and we applied it to the case of
HoNi$_2$B$_2$C. We interpreted the main anomaly
of its upper critical field via the
reduction of the interaction between phonons and
electrons in the Bloch-states of the magnetic structure. 
In this respect, the effect of a helical 
magnetic background on superconductivity is identical to
the effect of antiferromagnetism.
The helical case can be treated analytically
independently of the periodicity and the incommensurate
limit does not introduce additional suppression.
Finally, 
the anisotropy of the magnetic field and the temperature phase diagram
are well reproduced by taking into account
the magnetic response of the material.

We would like to thank H. Rosner for the data on the LDA
band structure of HoNi$_2$B$_2$C.
This work was performed under DFG Sonderforschungsbereich 463.

\bibliography{/home/amici/Bibliography/borocarbides,/home/amici/Bibliography/superconductivity}

\begin{thebibliography}{10}
\bibitem{Ho:mor96b}
A.~I. Morosov, Sov. Phys. JETP {\bf 83},  1048  (1996), note that this
  transcription of the author's name from Russian is different from the one
  used in the other references.
\end{thebibliography}

\end{document}